\input{psfig.sty}
      [1999/12/01 v1.4c Il Nuovo Cimento]
\documentclass{cimento}

\title{The short GRB 051210 observed by {\it Swift}}
\author{V.~La~Parola \from{ins:1}\ETC,
V.~Mangano\from{ins:2}, 
B.~Zhang\from{ins:3}, 
G.~Cusumano\from{ins:2},
T.~Mineo\from{ins:2}, 
E.~Troja\from{ins:2},
D.N.~Burrows\from{ins:4},
S.~Campana\from{ins:5}, 
M.~Capalbi\from{ins:6},
G.~Chincarini\from{ins:6},
P.~Giommi\from{ins:6},
A.~Moretti\from{ins:5},
M.~Perri\from{ins:6},
P.~Romano\from{ins:5},
G.~Tagliaferri\from{ins:6}.\\}
\instlist{
  \inst{ins:1} Universit\`a degli studi di Milano-Bicocca, Dipartimento di
               Fisica, Milan, Italy         	       
  \inst{ins:2} INAF- Istituto di Astrofisica Spaziale e Fisica Cosmica di Palermo
  \inst{ins:3} Department of Physics, University of Nevada, Las Vegas, USA
  \inst{ins:4} Department of Astronomy and Astrophysics, Pennsylvania State 
               University, USA     
  \inst{ins:5} INAF -- Osservatorio Astronomico di Brera, Merate, Italy
  \inst{ins:6} ASI Science Data Center, Frascati, Italy   
	       }
\PACSes{\PACSit{98.70.Rz}{$\gamma$-ray sources, $\gamma$-ray bursts.}
\PACSit{01.30.Cc}{Conference proceedings.}}
\begin{document}

\maketitle

\begin{abstract}
We report on the short GRB051210 detected by the {\it Swift}-BAT. The light 
curve, on which we focus mainly, shows a hint of extended emission in the BAT
energy range, a steep decay of the X-ray emission, without any flattening or
break, and two small flares in the first 300 sec. The emission fades out after
$\sim1000$ s. 
\end{abstract}

GRB051210 was classified as short after the onground analysis of the BAT data 
\cite{ref:bart} which revealed a
T$_{90}=1.27\pm0.05$ s. Moreover, both its spectral lag \cite{ref:bart} and its 
position on a T$_{90}$ vs. hardness plot are typical of short GRBs 
\cite{ref:kou}. 
No optical counterparts were identified by any ground based or space
telescope. A detailed analysis and discussion of the temporal and spectral
behaviour of this burst is in \cite{ref:lap}.

The rapid fading of the source and the lack of any flattening in the light 
curve after the initial decay may indicate that the GRB occurred in an 
extremely low density medium (naked GRB, \cite{ref:kum})  where the 
radiation emitted by the forward shock is 
expected to be undetectable. The steep decay of the X-ray emission is fully 
consistent with the hypothesis that we are observing a low energy tail of the 
prompt emission from an internal shock through the so-called curvature effect 
\cite{ref:zha}. Under this hypothesis
the emission would decay as $t^{-\alpha} = t^{(-\Gamma+1)}$ , where $\Gamma$ 
is the photon index of the GRB emission. In this case we get
$\Gamma + 1 = 2.54$, in very good agreement with the observed slope 
($2.58\pm0.11$). 
We can derive an estimation for the density of the interstellar medium {\it
n} in the vicinity of the burst from the expression of the 
expected afterglow flux according to the standard 
models \cite{ref:sa1}. Assuming
reasonable values for the redshift, the electron index and the total energy of
the burst, we get $n < 3 \times 10^3$ cm$^{-3}$, 
confirming the trend that short 
GRBs tend to be located in low-density  environments \cite{ref:sod}.

The extrapolation of the XRT light curve back to the burst onset does not 
match the BAT points by a few decades. This can be explained if the XRT 
light curve is the tail of a flare peaking at a time before the XRT 
observation and too weak to be detected by the BAT. Within such an 
interpretation, the zero time point of 
the rapid decay component should be shifted to the beginning of the rising 
segment of the relevant flare \cite{ref:zha}, which 
marks the reactivation of the central engine. The few marginally 
significant points in the BAT light curve starting at $\sim$T+20 s suggest that 
this could be the case of extended emission, as reported by \cite{ref:nor} for a 
few short bursts observed with BATSE. 

Two small flares are visible in the first 300 s after the onset in the XRT light
curve.
Delayed activity from the inner engine as an origin for flares (e.g. 
\cite{ref:bur}) can hardly be applied to short GRBs, if they originate in the 
merger of two compact objects in a binary system (NS-NS or NS-BH): 
hydrodynamical simulations suggest that the central engine activity of 
merger events cannot last more than a few seconds \cite{ref:dav}. 
Some alternative hypotheses: gravitational 
instability can lead to the fragmentation of the accretion 
disc, creating blobs of infalling material that produces 
the observed flares \cite{ref:per}. \cite{ref:prz} suggest that magnetic fields may 
build up near the black hole and form a magnetic barrier that can turn on and 
off the accretion episodes, leading to 
erratic X-ray flares at late epochs. 
\begin{figure}
\centerline{\psfig{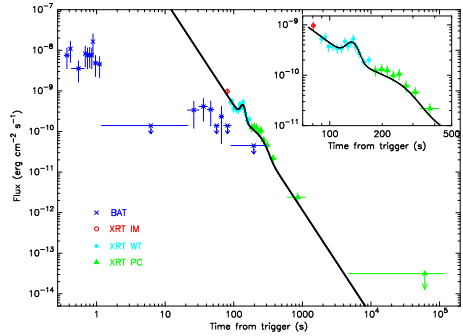}}
\caption{XRT light curve decay of GRB\,051210. The XRT and BAT count rates
were converted into flux units (0.2-10 keV) by applying a conversion factor 
derived from the relevant spectral analysis. The solid line represents the best 
fit model to the XRT data.}
\label{lc}
\end{figure}


\begin{thebibliography}{0}
\bibitem{ref:bart} \BY{Barthelmy et al.}, 2005, GCN 4321;
\bibitem{ref:kou} \BY{Kouveliotou et al.} \IN{ApJ L}{413}{1993}{101};
\bibitem{ref:lap} \BY{La Parola et al.} \IN{A\&A}{454}{2006}{753};
\bibitem{ref:kum} \BY{Kumar \atque Panaitescu} \IN{ApJ}{541}{2000}{51};
\bibitem{ref:zha} \BY{Zhang et al.} ApJ, 2006, in press;
\bibitem{ref:sa1} \BY{Sari \& Esin} \IN{ApJ}{548}{2001}{787};
\bibitem{ref:sod} \BY{Soderberg et al.} 2006, astroph/0601455;
\bibitem{ref:nor} \BY{Norris \atque Bonnell} ApJ, 2006, in press;
\bibitem{ref:bur} \BY{Burrows et al.} \IN{Science}{309}{2005}{1833};
\bibitem{ref:dav} \BY{Davies et al.} \IN{MNRAS}{356}{2005}{54};
\bibitem{ref:per} \BY{Perna et al.} \IN{ApJ L}{636}{2006}{29};
\bibitem{ref:prz} \BY{Proga \& Zhang} 2006, astroph/0601272

\end{thebibliography}
\end{document}